\documentclass[preprint2]{aastex} 
\begin{document}
\title {KECK-NIRSPEC INFRARED OH LINES: 
OXYGEN ABUNDANCES IN  METAL-POOR STARS DOWN TO [Fe/H] = $-$2.9
\footnote{Observations carried out with the Keck Telescope within
the Gemini-Keck agreement, and at the European Southern Observatory}}

\newcommand{\teff}{T$_{\rm eff}$ }
\newbox\grsign \setbox\grsign=\hbox{$>$} \newdimen\grdimen \grdimen=\ht\grsign
\newbox\simlessbox \newbox\simgreatbox
\setbox\simgreatbox=\hbox{\raise.5ex\hbox{$>$}\llap
     {\lower.5ex\hbox{$\sim$}}}\ht1=\grdimen\dp1=0pt
\setbox\simlessbox=\hbox{\raise.5ex\hbox{$<$}\llap
     {\lower.5ex\hbox{$\sim$}}}\ht2=\grdimen\dp2=0pt\def\simgreat{\mathrel{\copy\simgreatbox}}
\def\simless{\mathrel{\copy\simlessbox}}
\newbox\simppropto
\setbox\simppropto=\hbox{\raise.5ex\hbox{$\sim$}\llap
     {\lower.5ex\hbox{$\propto$}}}\ht2=\grdimen\dp2=0pt
\def\simpropto{\mathrel{\copy\simppropto}}

\author{\bf Jorge Mel\'endez
and Beatriz Barbuy}
\affil{Universidade de S\~ao Paulo, IAG, Dept. de Astronomia,
 CP 3386, S\~ao Paulo 01060-970, Brazil \\ E-mail:
jorge@astro.iag.usp.br, barbuy@astro.iag.usp.br}

\slugcomment{Submitted to the The Astrophysical Journal}
\slugcomment{Send proofs to:  J. Melendez}

\begin{abstract}

Infrared OH lines at 1.5 - 1.7 $\mu$m in the $H$ band were 
obtained with the NIRSPEC high-resolution spectrograph
at the 10m Keck Telescope for a sample of seven metal-poor stars.
Detailed analyses have been carried out, based on optical
high-resolution data obtained with the FEROS spectrograph at ESO.
Stellar parameters  were derived by adopting infrared flux method effective 
temperatures, trigonometric and/or evolutionary gravities and metallicities 
from \ion{Fe}{2} lines. We obtain that the sample stars with metallicities 
[Fe/H] $<$ $-$2.2 show a mean oxygen abundance [O/Fe] $\approx$ 0.54,
for a solar oxygen abundance of $\epsilon$(O) = $\approx$ 8.87, 
or [O/Fe] $\approx$ 0.64 if $\epsilon$(O) = 8.77 is assumed.

\end{abstract}

\section{Introduction}

Oxygen abundances in metal-poor stars are a key information for
understanding the early phases of Galactic chemical evolution.
Very few data on oxygen abundances are available for stars
with metallicities around [Fe/H] $\approx$ -3.0, and results
obtained from different lines are often discrepant.

The infrared (IR) X$^2 \Pi$ vibration-rotation transitions of OH lines
in the $H$ band were first used in halo stars by Balachandran \& Carney (1996)
for the dwarf  HD 103095, having derived [Fe/H] = $-$1.22 and [O/Fe] = 0.29.
Balachandran et al. (2001, 2002) and Mel\'endez et al. (2001)
presented new oxygen abundance determinations in metal-poor stars,
from OH lines in the $H$ band.  Their [O/Fe] values tend to show
agreement with those derived from [OI] lines
in the metallicity range -1.0 $\simless$ [Fe/H] $\simless$ -2.5;
three among  the four analyzed stars of lower metallicities give higher values.

The oxygen abundance determination using IR OH lines may solve the 
problem of disagreement between the values obtained from different 
oxygen abundance indicators.
The IR OH lines in the H band of the first overtone have measurable 
intensities down to [Fe/H] $\sim$ -3.0 for giants with effective temperatures
T$_{\rm eff}$ $\sim$ 4500 K,  and [Fe/H] $\sim$ -3.5 for T$_{\rm eff}$ $\sim$ 4300 K, 
whereas for dwarfs the lines are stronger, and with effective temperatures
T$_{\rm eff}$ $\sim$ 4500 K, stars of [Fe/H] $\sim$ -3.5 could be measured.
Otherwise only the fundamental transition lines at 3.5 $\mu$m, in the L band, 
are measurable. (The difficulty with these latter lines comes, however, from the
thermal background and lower sensitivity of the instruments.)
The relative intensity of the OH lines in the $L$ and $H$ bands
can be seen in Hinkle et al. (1995) for Arcturus, and for the Sun
in Livingston \& Wallace (1991), in which only the $L$-band lines are seen.

IR vibration-rotation OH lines tend to form in LTE, whereas the
electronic transition UV lines tend to form in non-LTE
 (Hinkle \& Lambert 1975). Non-LTE effects in the formation of OH will 
affect both sets of lines, according to 
Asplund \& Garc\'{\i}a-P\'erez (2001, hereafter AGP01).

In the present work we use IR OH lines in the region
1.5 - 1.7 $\mu$m in order to derive oxygen-to-iron ratios
for a sample of seven metal-poor stars, most of them 
in the metallicity range -3.0 $<$ [Fe/H] $<$ -2.0.

In Sect. 2 the observations are presented. In Sect. 3 the detailed
analyses are described. In Sect. 4 the results from IR OH lines 
are discussed, and in Sect. 5 conclusions are drawn.

\section{Observations}

\subsection{Data Acquisition and Reduction}

High-resolution infrared spectra were obtained from images taken  
at the 10m Keck Telescope, using the NIRSPEC spectrograph
(McLean et al. 1998). FWHM resolutions of 
37,000 were achieved with the \'echelle grating, 
with a 2-pixel slit width. The detector is an Aladdin 1024x1024 InSb array, 
covering essentially all the $H$ band in the range 1.5-1.7 
$\mu$m, apart from small gaps (0.008 $\micron$) between the orders.

Each star was observed at two slit positions with the same
integration time. A final flatfield and a dark image were prepared by
the Gemini/Keck team.

These infrared data were reduced using IRAF. The sky background 
was eliminated by subtracting the exposures from the sky obtained 
in the first/second exposures at the same position in the detector. 
Flat-field and dark corrections were applied. The orders were extracted 
with the task APALL. The wavelength calibration was carried out using 
the provided Ar-Ne-Xe-Kr lamp spectrum. Signal-to-noise ratios (S/Ns) of
$\approx$ 150-400 were estimated from continuum regions and reported in Table 1.

Optical spectra were obtained at the 1.52m telescope at ESO, La Silla, 
using the Fiber-fed Extended Range Optical Spectrograph (FEROS) 
(Kaufer et al. 2000). The total spectrum coverage is  356-920 nm 
with a resolving power of 48,000.  Two fibers, with entrance aperture 
of 2.7 arcsec, simultaneously recorded starlight and sky background. 
The detector is a back-illuminated CCD with 2948$\times$4096 pixels of 
15 $\mu$m size. Sample stars were reduced through a special pipeline package 
for reductions (DRS) of FEROS data, in a MIDAS environment. 
The data reduction proceeded with subtraction of bias and scattered light 
in the CCD, orders extraction, flat-fielding, and wavelength calibration with 
a ThAr calibration frame. 

The log of observations is given in Table 1. 

\section{DETAILED ANALYSIS}

The sample consists of six very metal poor giants and
one dwarf of intermediate metallicity.  

\subsection{Effective Temperatures}

Colors available in the literature were taken from the following sources:
$J$, $H$ and $K$ in the TCS (Telescopio Carlos S\'anchez) system
from Alonso, Arribas \& Mart\'{\i}nez-Roger (1994, 1998); Str\"omgren
{\it ubvy}-$\beta$ from the Catalogue by Hauck \& Mermilliod (1998); 
$V-R$ (Cousins) from McWilliam et al. (1995); and $V-R$,
$J$, $H$ and $K$  from the General Catalogue of Photometric Data 
(Mermilliod, Mermilliod \& Hauck 1997). Transformations between
different photometric systems were calculated using relations
given by Bessell (1979), Bessell \& Brett (1988) and 
Alonso et al. (1994, 1998).

The reddening values of most  stars, estimated  by using the
maps of reddening by Burstein \& Heiles (1982), were taken from 
Anthony-Twarog \& Twarog (1994). Otherwise, they were determined    
using maps by Arenou et al. (1992) and several high-Galactic latitude 
surveys, as implemented in a FORTRAN code by Hakkila et al. (1997).
The reddening $E(B-V)$ and the dereddened colors are given in
Table 2. Distances were determined from $Hipparcos$ parallaxes (Perryman 
et al. 1997) for 2 stars; otherwise we used the same procedure as
Anthony-Twarog \& Twarog (1994). 

Effective temperatures T$_{\rm eff}$ determined using the $b-y$, $V-R$, 
$V-K$ and $J-K$ calibrations of Alonso, Arribas \& Mart\'{\i}nez-Roger 
(1996a, 1999a,  hereafter AAM99a) are given in Table 3.
In this table are also shown infrared flux method (IRFM, column 7) 
temperatures, as determined by Alonso, Arribas \& Mart\'{\i}nez-Roger 
(1996b, 1999b, hereafter AAM99b). The mean of the IRFM calibrations 
(AAM99a,AAM99b), which are essentially the same as the IRFM
determinations by AAM99b, are shown in column 8 of Table 3.
These mean values were  checked against excitation equilibrium of \ion{Fe}{1} 
lines.  The temperatures based on excitation equilibrium of \ion{Fe}{1}  are
in the mean 18 K hotter than the mean of IRFM temperatures.
The small difference between \ion{Fe}{1} temperatures and the IRFM temperatures 
is an indication that non-LTE effects are minor (Tomkin \& Lambert 1999).

\subsection{Gravities}

Nissen et al. (1997) and  Allende Prieto et al. (1999) have shown 
that LTE spectroscopic gravities are lower than trigonometric gravities 
derived from $Hipparcos$ parallaxes in metal-poor stars. For this reason, 
$Hipparcos$ parallaxes $\pi$ were used to derive trigonometric gravities, 
when parallax values were available, which was the case for two stars. 
For the others, we derived the gravity by estimating their absolute magnitudes 
in colour-magnitude diagrams, using the calibrations by Norris et al. (1985).
These gravities are given in Table 4. The same method was also applied to the 
stars of Mel\'endez et al. (2001), since their parallaxes, and consequently their
gravities, had large uncertainties. The revised evolutionary gravities log $g$ 
and oxygen abundances are presented in Table 5.

Note that ionization equilibrium should not be used, given that
\ion{Fe}{1} lines show non-LTE effects (Th\'evenin \& Idiart 1999).

\subsection{Metallicities}

Equivalent widths of \ion{Fe}{1} and \ion{Fe}{2} lines for 5 stars 
(except for HD 134440 and HD 122956)were measured on high-resolution 
spectra obtained with the FEROS spectrograph at ESO.
For HD 134440, we adopted equivalent widths from Ryan \& Deliyannis (1998), 
Tomkin \& Lambert (1999), and Fulbright (2000), and for HD 122956, those 
by Sneden et al. (1991) and  Fulbright (2000).

We compare  the equivalent widths measured from the FEROS spectra with 
those from the Lick Observatory. The comparison of 20 lines from FEROS 
with those from the Hamilton/Lick data  by Sneden et al. (1991) gives
$\delta$(EW) = EW$_{ESO}$ - EW$_{Lick}$ = +0.1$\pm$0.7 m{\rm \AA} 
($\sigma$ = 3.2 m{\rm \AA}), and another 40 lines with Fulbright (2000) 
gives $\delta$(EW) = EW$_{ESO}$ - EW$_{Lick}$ = +0.8$\pm$0.5 m{\rm \AA} 
($\sigma$ = 2.9 m{\rm \AA}). 

MARCS model atmospheres by Bell et al. (1976) and Gustafsson et al. (1975)
were used for the calculations of curves of growth and spectrum synthesis.

The adopted oscillator strengths for the optical iron lines are from 
Nave et al. (1994), which correspond in most cases to $gf$-values
from O'Brian et al. (1991), and in cases where the lines are not given 
in O'Brian et al., the values are from the National Institute of Standards 
\& Technology (Fuhr, Martin, \& Wiese 1988) and a few other references.
The mean difference in log $gf$ values between O'Brian et al.(1991)
and the NIST database is of $\delta$log $gf$ (O'Brian et al - NIST) = 0.02.

On the other hand, $gf$-values of \ion{Fe}{2} lines show a larger variation 
in the literature. We have adopted theoretical $gf$-values by 
Bi\'emont et al. (1991), calibrated with experimental data as follows: 
lifetimes of upper levels from Schnabel et al. (1999), Schnabel \& Kock (2001), 
Guo et al. (1992), Hannaford et al. (1992), and Bi\'emont et al. (1991) 
and branching ratios by Heise \& Kock (1990), Pauls et al. (1990), and 
Kroll \& Kock (1987) are adopted, and $gf$-values are computed. These values 
are gathered in multiplets, and corrections are applied from comparisons between 
the theoretical and laboratory values.

Curves of growth and abundances from equivalent widths of \ion{Fe}{1} 
and \ion{Fe}{2} were computed using the codes RENOIR and ABON by M. Spite. 
Microturbulence velocities $v_t$ were obtained from  curves of growth
of \ion{Fe}{1}, and these values were checked by requiring no dependence 
of [Fe/H] against reduced equivalent width. 

In Table 6 are given the Fe abundances derived from  IRFM effective 
temperatures, trigonometric or evolutionary gravities, and curves of
growth based on optical \ion{Fe}{2} lines, and literature parameters 
for comparison purposes.

\subsection{Oxygen Abundances}

The oxygen abundances were determined from fits of synthetic spectra
to the sample spectra. The LTE code for spectrum synthesis described in 
Cayrel et al. (1991) was employed for the calculations. In all cases, 
the [O/Fe] value was obtained by adopting IRFM temperatures, trigonometric
gravities, and metallicities from  \ion{Fe}{2} lines (see discussions in 
Th\'evenin \& Idiart 1999; AGP01).

The list of  atomic lines present in the H band compiled by 
Mel\'endez \& Barbuy (1999) and Mel\'endez (1999) was adopted, together 
with molecular lines of CN A$^2$$\Pi$ - X$^2$$\Sigma$, CO X$^1$$\Sigma^+$ 
(Mel\'endez \& Barbuy 1999), and OH (X$^2$$\Pi$). The IR OH vibration-rotation 
lines (X$^2 \Pi$) used in this work, from the first-overtone ($\Delta v$ = 2) 
sequence, were described in Mel\'endez et al. (2001). Energy levels for the 
OH lines were computed from molecular parameters given in Co\-xon \& Foster (1992) 
and Abrams et al. (1994). Molecular oscillator strengths were calculated from 
Einstein coefficients given  by Goldman et al. (1998), which are  accurate 
to 10-15\%, depending on the molecular quantum numbers, according to the authors.

The list of OH lines used for oxygen abundance determination
together with their molecular $gf$-values, energy levels, and 
equivalent widths are given in Table 7. The fit of synthetic
spectra of OH lines for HD 110184 is shown in Figure 1.

The oxygen abundances derived show a dependence on the carbon abundances 
adopted. Carbon abundances were derived from the CH A$^2\Pi$-X$^2\Sigma$
$G$ band at $\sim$ 4300 {\rm \AA}, from optical FEROS spectra, 
for five stars. Details on the atomic and molecular database used are described 
in Castilho et al. (1999). The observed and synthetic spectra of the
$G$ band for HD 110184 are shown in Figure 2. For HD 122956, the 1.56$\mu$m CO 
lines were used to derive its carbon abundance, whereas for HD 134440, the value 
given in Carbon et al. (1987) was adopted.

The carbon and oxygen abundances derived from  IRFM T$_{\rm eff}$, 
trigonometric gravities, and [\ion{Fe}{2}/H]  are given in Table 8.

The solar oxygen abundance has been recently revised by Holweger (2001)
and Allende Prieto et al. (2001). The main change in the solar oxygen
abundance derived from the forbidden line is due to the presence of 
a \ion{Ni}{1} line not considered before, and $\epsilon$(O) = 8.74 was found 
by Allende Prieto et al. (2001). Holweger (2001) gives $\epsilon$(O) = 8.69 
using several permitted lines. Note that the oscillator strength of the 
[\ion{O}{1}] line adopted by Allende Prieto et al. (2001) is log $gf$ = -9.72, 
0.03 dex higher than previous values around log $gf$ = -9.75, leading to lower
oxygen abundances. Grevesse et al. (1996) reported $\epsilon$(O) = 8.87, whereas
Grevesse \& Sauval (1998) give $\epsilon$(O) = 8.83. The value from 
Allende Prieto et al. (2001) is obtained with three-dimensional hydrodynamical 
model atmospheres, whereas if one-dimensional hydrostatic model atmospheres are
used, a value of $\epsilon$(O) = 8.77 is obtained. We adopted $\epsilon$(O) = 8.87, 
and if  $\epsilon$(O) = 8.77 is to be considered, the oxygen abundances will be  
0.1 dex higher.

\subsection{Errors}

The errors due to uncertainties on T$_{\rm eff}$, log $g$, and $v_t$ are 
inspected by computing the results for HD 110184. In Table 9 the errors 
for $\Delta$T$_{\rm eff}$ = 100 K, $\Delta$log $g$ = 0.5 dex, 
and $\Delta v_t$ = 0.5 km s$^{-1}$ are given.

\section{Discussion}

\subsection{Reliability of the IR OH lines}

OH is a trace component of oxygen abundances in stellar atmospheres, 
and its measurement is very susceptible to errors in the adopted temperature 
structure of model atmospheres (Asplund 2001, 2002; AGP01; Lambert 2002).
Using 3D hydrodynamical model atmospheres, AGP01 have demonstrated the
extreme sensitivity of molecule formation and derivation of oxygen abundances 
in metal-poor stars relative to calculations with classical 1D hydrostatic 
model atmospheres and have shown that the oxygen abundance using the 
classical models  may  overestimate the derived oxygen abundance.

According to AGP01, for solar or moderate metallicities ([Fe/H] $>$ -1.0), 
the temperature remains close to the radiative equilibrium value; however,
at lower metallicities, the temperature in the outer layers departs
significantly from it (Asplund et al. 1999). Radiative heating occurs due 
to spectral line absorption of continuum photons coming from deeper layers. 
In metal-poor stars, adiabatic cooling dominates over radiative heating, and 
the effects of lower temperatures are first seen at the outermost layers, moving 
toward deeper layers for the lower metallicities. At log $\tau_{500nm}$
$\sim$ -3, the average temperature difference between 3D hydrodynamical and 
1D hydrostatic models can reach 1000 K (AGP01).

AGP01 verified that OH lines in the IR, with similar line strengths
as the UV lines, suffer from granulation abundance corrections just as severe.
In Table 9 we give the equivalent widths (EWs) of the different OH lines 
measured: there is a clear increase in the oxygen abundance derived with 
increasing EW, as shown in Figure 3. Given that the lower EW lines are more 
reliable, according to AGP01, we adopt an oxygen abundance resulting from a 
mean of the five weaker OH lines. It is interesting to note that the effect is 
stronger for the most metal-poor stars.

\subsection{Comparison with the literature}

In Figure 4 the  [O/Fe] versus [Fe/H] obtained from IR OH lines in the 
present work are plotted.

This result is compared with results from the forbidden doublet of [\ion{O}{1}]$\lambda$6300.31, $\lambda$6363.79 {\rm \AA} lines, 
which is the most reliable oxygen abundance indicator, given that
it is immune to non-LTE effects (Kiselman 2001, 2002; Lambert 2002),
and besides neutral oxygen is the dominant species in stellar atmospheres.
For this reason, we compare the present results based on IR OH lines 
with results in the literature obtained from the [\ion{O}{1}] lines
(e.g. Gratton \& Ortolani 1986; Barbuy 1988; Barbuy \& Erdelyi-Mendes 1989; 
Sneden et al. 1991; Spite \& Spite 1991;  Kraft et al. 1992; 
Nissen \& Edvardsson 1992; Shetrone 1996; Balachandran \& Carney 1996; 
Fulbright \& Kraft 1999; Gratton et al. 2000; Westin et al. 2000;
Balachandran et al. 2001, 2002; Cayrel et al. 2001; Nissen et al. 2001;
Smith et al. 2001; Sneden \& Primas 2001, 2002; Cowan et al. 2002).
(Note that in Mel\'endez et al. 2001, we have discussed oxygen abundance 
results from IR OH lines, as compared to UV OH lines and \ion{O}{1} lines, 
and we defer the reader to that discussion for those comparisons.)

The results from the [\ion{O}{1}] line were selected according to
their S/Ns, such that only results in which the EW is larger than 5 times
the error were considered, and we used Cayrel's (1988) formula for the 
error on the EW 
$\delta W_{\lambda}$$\approx$1.6(w$\times$$\delta x$)$^{1/2}$/\-(S/N),
where w is the FWHM of a typical spectral line and $\delta x$ is the
pixel size. In Figs. 5a-c we plot [O/Fe] versus [Fe/H] from the 
references above. In Figure 5a  the results from different authors are 
indicated; for 26 stars, two or more different results are plotted.
In Figure 5b giants (log $g$ $<$ 2.5, $open circles$) and dwarfs 
(log $g$ $>$ 2.5, $filled circles$) are indicated, where the mean of 
values for the same star are considered, differently from Figure 5a, 
in which all values reported in the listed references are plotted. 
In Figure 5c the means of [O/Fe] values, as given in Table 10,
are plotted in bins of 0.2 dex in [Fe/H], and the size of the circles 
represents the number of stars in each bin. 

Oxygen abundances from IR OH lines at metallicities [Fe/H] $<$ -2.2 
show a trend to higher values, in agreement with chemical evolution models. 
The behavior is also similar to that shown by Sneden \& Primas (2001, 2002)
for [O/Sc] vs. [Fe/H]. Finally, it is important to note that
mixing along the red giant branch could deplete the oxygen
in giants, but the effect should be minor, as studied by
Gratton et al. (2000).

\section{Conclusions}

We obtained high-resolution infrared spectra in the $H$ band in order 
to derive oxygen abundances from IR OH lines. In order to have a homogeneous 
set of stellar parameters, we carried out detailed analyses using equivalent 
widths of iron lines measured on high-resolution spectra from the FEROS 
spectrograph at ESO.

\begin{enumerate}
\item {The sample stars with metallicities [Fe/H] $<$ $-$2.2 show a slight 
increase in [O/Fe], with respect to the results for higher metallicity stars.
A mean of [O/Fe] = 0.5 is found, similar to the value derived from the 
forbidden [OI] lines.}

\item{
A clear increase in the oxygen abundance derived with increasing equivalent 
width is found, such that the stronger lines give higher abundances.
We adopt an oxygen abundance resulting from a mean of the weaker OH lines.
It is also important to note that the effect is more pronounced for the 
most metal-poor stars.}

\item{ It is clear that oxygen abundance determinations for the more
metal-poor stars derived from very high S/N forbidden [\ion{O}{1}] and 
IR OH lines are needed, as well as 3D model atmospheres calculations, in the 
case of IR OH lines.}

\end{enumerate}

\acknowledgements
We acknowledge partial financial support from FAPESP and CNPq.
J.M. acknowledges the FAPESP postdoctoral fellowship 01/01134-3.
We  have made use of data from the $Hipparcos$ astrometric mission of the ESA.
This research is based on observations obtained by staff of the Gemini Observatory, 
which is operated by the Association of Universities for Research in Astronomy, Inc.,
under a cooperative agreement with the NSF on behalf of the Gemini partnership: 
The National Science Foundation (United States), the Particle Physics and 
Astronomy Research Council (United Kingdom), the National Research Council (Canada),
CONICYT (Chile), the Australian Research Council (Australia), CNPq (Brazil), and
CONYCET (Argentina). The IR data presented herein were obtained at the W.M. Keck
Observatory, which is operated as a scientific partnership among 
the California Institute of Technology, the University of California, and 
the National Aeronautics and Space Administration. The observatory was made 
possible by the generous financial support of the W.M. Keck Foundation.
The FEROS observations at the European Southern Observatory (ESO) were
carried out within the Observat\'orio Nacional ON/ESO and
ON/IAG agreements, under Fapesp project 1998/10138-8.

\begin{deluxetable}{llllllll}
\label{observations}
\tablewidth{0pt}
\footnotesize
\tablecaption{Log of observations}
\tablehead{
\colhead{Star}  &
\colhead{$V$}      &
\colhead{Date }      &
\colhead{Exposure (s)}      &
\colhead{S/N} &
\colhead{Wavelength} &
\colhead{Telescope} }
\startdata
HD 2796   & 8.50 & 2001-11-20 & 12$\times$300 & 150 & 1.49-1.75$\mu$m & Keck \\
HD 6268   & 8.11 & 2000-12-15 & 12$\times$100 & 450 & 1.49-1.75$\mu$m & Keck \\
BD-180271 & 9.85 & 2000-12-15 & 13$\times$200 & 400 & 1.49-1.75$\mu$m & Keck \\
HD 110184 & 8.31 & 2000-05-20 & 8$\times$80   & 180 & 1.46-1.70$\mu$m & Keck \\
HD 122956 & 7.22 & 2000-05-20 & 6$\times$60   & 300 & 1.46-1.70$\mu$m & Keck \\
HD 134440 & 9.44 & 2000-05-20 & 8$\times$150  & 200 & 1.46-1.70$\mu$m & Keck \\
BD-185550 & 9.26 & 2001-06-12 & 12$\times$120 & 350 & 1.49-1.75$\mu$m & Keck \\
HD 2796   & 8.50 & 2001-10-01 & 2400 & 240 & 360-900nm & ESO \\
HD 6268   & 8.11 & 2001-10-01 & 1800 & 320 & 360-900nm & ESO \\
BD-180271 & 9.85 & 2001-10-01 & 3600 & 340 & 360-900nm & ESO \\
HD 110184 & 8.31 & 2001-01-18 & 1800 & 300 & 360-900nm & ESO \\
BD-185550 & 9.26 & 2001-10-01 & 3600 & 200 & 360-900nm & ESO \\
\enddata
\end{deluxetable}

\begin{deluxetable}{lllllll}
\label{colors}
\tablewidth{0pt}
\footnotesize
\tablecaption{Colors}
\tablehead{
\colhead{Star} &
\colhead{E(B-V)} &
\colhead{{(b-y)}$_{\rm 0}$} &
\colhead{{(V-R)}$^{\rm J}_{\rm 0}$} &
\colhead{{(J-K)}$^{\rm TCS}_{\rm 0}$} &
\colhead{{(V-K)}$^{\rm TCS}_{\rm 0}$}
}
\startdata
HD 2796   & 0.068 & 0.538 & 0.714 & 0.540 & 2.212 \\
HD 6268   & 0.016 & 0.586 & 0.770 & ----  &  ---- \\
BD-180271 & 0.005 & 0.841 & 1.026 & 0.728 & 3.010 \\
HD 110184 & 0.0 & 0.818 & 0.977 & 0.702 & 2.918 \\
HD 122956 & 0.041 & 0.638 & 0.813 & ----  & 2.558 \\
HD 134440 & 0.0 & 0.522 & 0.762 & 0.586 & 2.293 \\
BD-185550 & 0.117 & 0.599 & 0.791 & 0.598 & 2.389 \\
\enddata
\tablecomments{Subscript J is for Johnson; subscript TCS is for 
Telescopio Carlos S\'anchez}
\end{deluxetable}

\begin{deluxetable}{llllllllll}
\tablecolumns{8}
\tablewidth{0pt}
\footnotesize
\tablecaption{Effective temperatures (K) based on IRFM (Sect. 3.1)}
\tablehead{
\multispan1 {Star} & 
\multispan5 {IRFM calibration (AAM99a)} & 
\multispan1 {IRFM} & 
\multispan1 {IRFM} \\
\cline{2-6}
\colhead{}   &
\colhead{b-y} &
\colhead{V-R} &
\colhead{V-K} &
\colhead{J-K} &
\colhead{mean} &
\colhead{AAM99b} &
\colhead{mean} 
}
\startdata
HD 2796   & 4840 & 4822 & 4858 & 4882  & 4851 & 4867 & 4859  \\
HD 6268   & 4739 & 4670 & ---- & ----  & 4705 & ---  & 4705  \\
BD-180271 & 4239 & 4112 & 4190 & 4277  & 4212 & 4277 & 4245  \\
HD 110184 & 4272 & 4199 & 4249 & 4376  & 4274 & 4250 & 4262  \\
HD 122956 & 4553 & 4547 & 4519 & ----  & 4540 &  --- & 4540  \\
HD 134440 & 4850 & 4638 & 4709 & 4669  & 4717 & 4746 & 4732  \\
BD-185550 & 4789 & 4637 & 4679 & 4682  & 4697 & 4668 & 4683  \\
\enddata
\label{teffs}
\end{deluxetable}

\begin{deluxetable}{llllll}
\tablewidth{0pt}
\footnotesize
\tablecaption{Gravities (log {\it g})}
\tablehead{
\colhead{Star}   &
\colhead{$Hipparcos$} &
\colhead{Evolutionary}  &
\colhead{Mean (Adopted)}
}
\startdata
HD 2796   &    ---       & 1.7 & 1.7$\pm$0.3  \\
HD 6268   &    ---       & 1.5 & 1.5$\pm$0.3  \\
BD-180271 &    ---       & 0.7 & 0.7$\pm$0.3  \\
HD 110184 &      ---     & 0.7 & 0.7$\pm$0.3  \\
HD 122956 & 1.7$\pm$0.3  & 1.5 & 1.6$\pm$0.3  \\
HD 134440 & 4.8$\pm$0.04 & 4.8 & 4.8$\pm$0.1  \\
BD-185550 &       ---    & 1.7 & 1.7$\pm$0.3  \\
\enddata
\label{logg}
\end{deluxetable}

\begin{deluxetable}{lllll}
\tablewidth{0pt}
\footnotesize
\tablecaption{Revised [O/Fe] values from OH-IR lines}
\tablehead{
\colhead{Star}  &
\colhead{\teff$^{\rm IRFM}$} &
\colhead{log $g^{\rm Hip}$}  &
\colhead{[\ion{Fe}{2}/H]} &
\colhead{[O/Fe]} }
\startdata
HD 2665    & 4980 & 2.3  & $-$1.90 & 0.48 \\
HD 6582    & 5305 & 4.5  & $-$0.83 & 0.26 \\
HD 6755    & 5030 & 2.9  & $-$1.56 & 0.16 \\
HD 21581   & 4870 & 2.7  & $-$1.50 & 0.20 \\
HD 25329   & 4810 & 4.8  & $-$1.76 & 0.53 \\
HD 26297   & 4320 & 1.1  & $-$1.64 & 0.27 \\
HD 29574   & 4030 & 0.5  & $-$1.70 & 0.15 \\
HD 37828   & 4390 & 1.6  & $-$1.28 & 0.43 \\
HD 103095  & 5035 & 4.7  & $-$1.27 & 0.25 \\
HD 165195  & 4240 & 0.7  & $-$2.14 & 0.43 \\
HD 187111  & 4260 & 0.7  & $-$1.67 & 0.51 \\
HD 206739  & 4650 & 1.7  & $-$1.52 & 0.29 \\
HD 216143  & 4490 & 1.0  & $-$2.08 & 0.57 \\
HD 221170  & 4410 & 1.1  & $-$2.00 & 0.21 \\
BD +060648 & 4220 & 0.8 & $-$1.89 & 0.30 \\
\enddata
\label{tabxxxx}
\end{deluxetable}

\begin{deluxetable}{lllllllllllllll}
\tablecolumns{15}
\tablewidth{0pt}
\scriptsize
\tablecaption{Adopted atmospheric parameters and literature values}
\tablehead{
\colhead{Star} &
\multicolumn{4}{c}{ IRFM, log $g$ and \ion{Fe}{2}} & 
\colhead{} &
\multicolumn{5}{c}{ Literature parameters (mean values)} \\[4pt]
\cline{2-5} \cline{7-11} \\
\colhead{}  &
\colhead{\teff$^{\rm IRFM}$} &
\colhead{log $g$}  &
\colhead{[\ion{Fe}{2}/H]} &
\colhead{v$_t$} &
\colhead{}  &
\colhead{\teff} &
\colhead{log $g$}  &
\colhead{[Fe/H]} &
\colhead{v$_t$} &
\colhead{ref.} 
}
\startdata
HD 2796   & 4860 &1.7 & $-$2.26 & 1.9 & & 
             4910 & 1.4 & $-$2.34 & 2.1 & 1,2,3,4 \\
HD 6268   & 4705 &1.5 & $-$2.35 & 1.9 & &
             4690 & 1.3 & $-$2.43 & 1.9 & 1,3,4 \\
BD-180271  & 4245 &0.7 & $-$2.35 & 2.5 & &
             4230 & 0.5 & $-$2.39 & 1.9 & 5 \\
HD 110184  & 4262 &0.7 & $-$2.41 & 2.4 & &
             4294 & 0.5 & $-$2.50 & 2.1 & 3,4,6,7 \\ 
HD 122956  & 4540 &1.6 & $-$1.60 & 1.6 & &
             4631 & 1.4 & $-$1.75 & 1.7 & 3,4,6-11 \\
HD 134440  & 4732 &4.8 & $-$1.44 & 1.0 & &
             4804 & 4.6 & $-$1.47 & 1.1 & 12-15 \\
BD-185550  & 4683 &1.7 & $-$2.87 & 1.7 & &
             4698 & 1.3 & $-$2.93 & 2.0 & 1,3,4,16-17 \\
\enddata
\label{tab4}
\tablerefs{(1) McWilliam et al. 1995; (2) Fran\c cois 1996; 
(3) Pilachowski et al. 1996; 
(4) Burris et al. 2000; (5) Shetrone 1996; (6) Sneden et al. 1991; 
(7) Fulbright 2000; 
(8) Gratton \& Sneden 1991; (9) Fran\c cois, Spite \& Spite 1993; 
(10) Gratton \& Sneden 1994;
(11) Gratton et al. 2000; (12) Tomkin \& Lambert 1999; 
(13) Ryan \& Deliyannis 1998; (14) King 1997; (15) Spiesman \& Wallerstein 1991
(16) Cavallo, Pilachowski \& Rebolo 1997
(17) Peterson, Kurucz, Carney 1990
}
\end{deluxetable}

\begin{deluxetable}{lllllllllllllllllll}
\tablewidth{0pt}
\tablecaption{Equivalent widths (m{\rm \AA}) of IR OH lines}
\scriptsize
\tablehead{
\colhead{v'-v", branch, J"}  &
\colhead{{$\lambda$}}   &
\colhead{{$\chi_{exc}$}}  &
\colhead{log {\it gf}} &
\colhead{HD} &
\colhead{HD} &
\colhead{BD} &
\colhead{HD} &
\colhead{HD} &
\colhead{HD} &
\colhead{BD} \\
\colhead{}  & 
\colhead{({\rm \AA})}   &
\colhead{(eV)}  &
\colhead{} &
\colhead{2796} &
\colhead{6268} &
\colhead{-1802} &
\colhead{1101} &
\colhead{1229} &
\colhead{1344} &
\colhead{-1855}
 }
\startdata
2-0 P2f  6.5  & 15002.15 & 0.134 & $-$5.578 & ... & ... & ... & ... & 19  & ... & ... \\
2-0 P2e  6.5  & 15003.12 & 0.134 & $-$5.578 & ... & ... & ... & ... & 21  & ... & ... \\
2-0 P1e  7.5  & 15021.04 & 0.127 & $-$5.520 & ... & ... &  48 &  51 & ... & ... & ... \\
2-0 P1f  7.5  & 15022.86 & 0.128 & $-$5.520 & ... & ... &  48 & ... & ... & ... & ... \\
2-0 P2f  8.5  & 15264.61 & 0.210 & $-$5.429 & ... & ... &  44 &  32 & ... & ... & ... \\
2-0 P2e  8.5  & 15266.17 & 0.210 & $-$5.429 & ... & ... &  42 &  31 & 23  & ... & 7.5 \\
2-0 P1e 9.5   & 15278.52 & 0.205 & $-$5.382 & ... & ... &  50 &  40 & 26  & ... & 8.0 \\
2-0 P1f 9.5   & 15281.05 & 0.205 & $-$5.382 & ... & ... &  49 &  39 & 26  & ... & 6.0 \\
3-1 P2e\&f 2.5& 15283.62 & 0.478 & $-$5.312 & ... & ... &  26 & ... & 16  & ... & ... \\
3-1 P2e\&f 3.5& 15391.14 & 0.494 & $-$5.137 & ... & ... &  36 & ... & ... & ... & ... \\
2-0 P2e  9.5  & 15407.29 & 0.255 & $-$5.365 & ... & ... & ... & ... & ... & ... & 4.5 \\
2-0 P2e  9.5  & 15409.17 & 0.255 & $-$5.365 & ... & ... &  48 & ... & ... & ... & 4.3 \\
2-0 P1e 10.5  & 15419.46 & 0.250 & $-$5.323 & ... & ... &  58 & ... & ... & ... & 7.5 \\
2-0 P1f 10.5  & 15422.37 & 0.250 & $-$5.323 & ... & ... &  58 & ... & ... & ... & 8.0 \\
3-1 P1e 5.5   & 15535.46 & 0.507 & $-$5.230 & ... & ... &  26 &  18 & 16  & 13  & 5.0 \\ 
3-1 P1f 5.5   & 15536.71 & 0.507 & $-$5.230 & ... & ... &  22 &  16 & 16  & 12  & 5.5 \\
2-0 P2e 10.5  & 15560.24 & 0.304 & $-$5.307 & ... & 12  &  52 &  35 & 31  & 23  & 7.0 \\
4-2 R2e\&f 2.5& 15565.91 & 0.899 & $-$5.003 & ... & ... &  16 & ... & ... & 10.5& ... \\
2-0 P1e 11.5  & 15568.78 & 0.299 & $-$5.269 & ... & 10  &  47 &  36 & 27  & 25  & 5.0 \\
2-0 P1f 11.5  & 15572.08 & 0.300 & $-$5.269 & ... & 11  &  52 &  38 & 28  & 29  & 7.2 \\
3-1 P2f  5.5  & 15626.70 & 0.541 & $-$5.198 & ... & ... & ... & ... & 21  & ... & ... \\
3-1 P2e  5.5  & 15627.41 & 0.541 & $-$5.198 & ... & ... & ... & ... & 22  & ... & ... \\
3-1 P1e  6.5  & 15651.90 & 0.534 & $-$5.132 & ... & ... &  30 &  22 & ... & ... & 5.5 \\
3-1 P1f  6.5  & 15653.48 & 0.534 & $-$5.133 & ... & ... &  34 &  24 & ... & ... & 5.5 \\
2-0 P2f 11.5  & 15717.12 & 0.357 & $-$5.254 & ... & ... &  55 & ... & ... & ... & ... \\
2-0 P2e 11.5  & 15719.70 & 0.358 & $-$5.254 & ... & ... &  52 & ... & ... & ... & ... \\
2-0 P1e 12.5  & 15726.72 & 0.353 & $-$5.219 & 7   & ... &  47 & ... & ... & ... & ... \\
2-0 P1f 12.5  & 15730.44 & 0.354 & $-$5.219 & 6   & ... &  45 & ... & ... & ... & ... \\
2-0 P1f 13.5  & 15897.70 & 0.412 & $-$5.172 & 7   & ... &  51 &  42 & ... & 40  & ... \\
3-1 P1e  8.5  & 15910.42 & 0.600 & $-$4.976 & ... & 7.3 &  50 &  40 & ... & ... & ... \\
3-1 P1f  8.5  & 15912.73 & 0.600 & $-$4.976 & ... & 6.3 &  47 &  35 & ... & 30  & ... \\
3-1 P2f  8.5  & 16036.89 & 0.645 & $-$4.957 & ... & ... &  37 & ... & ... & ... & ... \\
3-1 P2e  8.5  & 16038.54 & 0.645 & $-$4.957 & ... & ... &  48 & ... & ... & ... & ... \\
2-0 P2f 13.5  & 16061.70 & 0.476 & $-$5.159 & ... & ... &  43 & ... & ... & ... & 7.5 \\
2-0 P2e 13.5  & 16065.06 & 0.477 & $-$5.159 & ... & ... &  40 & ... & ... & ... & ... \\
3-1 P2e  9.5  & 16192.13 & 0.688 & $-$4.893 & 6   & ... & ... &  32 & 23  & 26  & 5.1 \\
3-1 P1e 10.5  & 16204.08 & 0.683 & $-$4.851 & ... & ... &  58 & ... & ... & ... & ... \\
3-1 P1f 10.5  & 16207.19 & 0.683 & $-$4.851 & ... & ... &  56 & ... & ... & ... & ... \\
2-0 P2f 14.5  & 16247.88 & 0.542 & $-$5.115 & ... & ... &  40 &  27 & 25  & 24  & 4.3 \\
2-0 P2e 14.5  & 16251.66 & 0.543 & $-$5.115 & ... & ... &  44 &  25 & 24  & 25  & 4.1 \\
2-0 P1e 15.5  & 16255.02 & 0.538 & $-$5.087 & ... & 6.6 &  49 &  28 & ... & 23  & ... \\  
2-0 P1f 15.5  & 16260.16 & 0.538 & $-$5.087 & ... & 7.0 &  48 &  28 & 26  & 25  & ... \\
4-2 P1e  5.5  & 16346.19 & 0.927 & $-$4.938 & ... & ... &  13 & ... & ... & ... & ... \\
4-2 P1f  5.5  & 16347.49 & 0.927 & $-$4.938 & ... & ... &  12 & ... & ... & ... & ... \\
3-1 P2f 10.5  & 16352.22 & 0.735 & $-$4.835 & ... & 6.0 &  54 & ... & ... & ... & 4.2 \\
3-1 P2e 10.5  & 16354.58 & 0.735 & $-$4.835 & ... & 5.9 &  51 & ... & ... & ... & 3.8 \\
3-1 P1f 11.5  & 16368.13 & 0.731 & $-$4.797 & ... & ... &  46 & ... & ... & ... & 4.0 \\
2-0 P2e 15.5  & 16448.06 & 0.612 & $-$5.075 & ... & ... & ... & ... & ... & 29  & ... \\
2-0 P1e 16.5  & 16450.37 & 0.608 & $-$5.048 & ... & ... & ... & ... & ... & 27  & ... \\
2-0 P1f  16.5 & 16456.04 & 0.609 & $-$5.048 & ... & ... & ... & ... & 25  & 26  & ... \\
4-2 P1f   6.5 & 16472.82 & 0.953 & $-$4.840 & ... & ... & ... & ... & 10.5& ... & ... \\
3-1 P2e 11.5  & 16526.25 & 0.787 & $-$4.782 & ... & ... &  36 &  44 & 21  & 29  & ... \\
3-1 P1e 12.5  & 16534.58 & 0.782 & $-$4.746 & ... & ... &  36 &  50 & ... & 26  & 4.6 \\
3-1 P1f 12.5  & 16538.59 & 0.783 & $-$4.746 & ... & ... &  35 &  42 & ... & ... & ... \\
4-2 P2f  6.5  & 16581.27 & 0.989 & $-$4.813 & ... & ... &  18 & ... & ... & 14  & ... \\
4-2 P2e  6.5  & 16582.32 & 0.989 & $-$4.813 & ... & ... &  22 & ... & ... & 13  & ... \\
4-2 P1e  7.5  & 16605.46 & 0.982 & $-$4.755 & ... & ... &  26 & ... & ... & 25  & ... \\
4-2 P1f  7.5  & 16607.52 & 0.982 & $-$4.755 & ... & ... &  20 & ... & 19  & 24  & ... \\
2-0 P2f 16.5  & 16649.95 & 0.685 & $-$5.036 & ... & ... &  24 &  31 & ... & ... & ... \\
2-0 P2e 16.5  & 16654.65 & 0.686 & $-$5.036 & ... & ... &  25 &  34 & ... & ... & ... \\
2-0 P1e 17.5  & 16655.99 & 0.682 & $-$5.010 & ... & 5.4 &  32 &  31 & ... & ... & ... \\
2-0 P1f 17.5  & 16662.21 & 0.683 & $-$5.011 & ... & 6.0 & ... &  31 & ... & ... & ... \\
3-1 P2f 12.5  & 16704.36 & 0.842 & $-$4.732 & ... & ... &  44 & ... & ... & ... & ... \\
3-1 P2e 12.5  & 16707.52 & 0.842 & $-$4.732 & ... & 6.5 &  43 & ... & ... & ... & ... \\
3-1 P1e 13.5  & 16714.36 & 0.837 & $-$4.698 & ... & 5.8 &  47 & ... & ... & ... & ... \\
4-2 P1f  8.5  & 16751.72 & 1.016 & $-$4.682 & ... & ... &  23 & ... & ... & ... & ... \\
3-1 P2f 13.5  & 16895.18 & 0.901 & $-$4.685 & ... & ... &  41 & ... & ... & ... & ... \\
3-1 P2e 13.5  & 16898.78 & 0.901 & $-$4.685 & ... & 6.0 &  39 & ... & ... & ... & ... \\
4-2 P1e  9.5  & 16902.73 & 1.054 & $-$4.616 & ... & ... &  26 & ... & ... & ... & ... \\
3-1 P1e 14.5  & 16904.28 & 0.897 & $-$4.654 & ... & 5.8 &  26 & ... & ... & ... & ... \\
4-2 P1f  9.5  & 16905.63 & 1.054 & $-$4.616 & ... & ... &  20 & ... & ... & ... & ... \\
3-1 P1f 14.5  & 16909.29 & 0.898 & $-$4.654 & ... & ... &  37 & ... & 24  & ... & ... \\
4-2 P1e 10.5  & 17066.13 & 1.095 & $-$4.556 & ... & ... &  21 & ... & ... & ... & ... \\
4-2 P1f 10.5  & 17069.48 & 1.096 & $-$4.556 & ... & ... &  19 & ... & ... & ... & ... \\
3-1 P1e 15.5  & 17104.72 & 0.960 & $-$4.613 & ... & 5.9 & ... & ... & ... & ... & ... \\
3-1 P2f 15.5  & 17308.44 & 1.031 & $-$4.600 & ... & ... &  20 & ... & ... & ... & ... \\
3-1 P2e 15.5  & 17312.99 & 1.032 & $-$4.600 & ... & ... &  22 & ... & ... & ... & ... \\
\enddata
\label{tabewoh}
\end{deluxetable}

\begin{deluxetable}{lllllllllllllll}
\tablecolumns{12}
\tablewidth{0pt}
\scriptsize
\tablecaption{Iron, Carbon, and Oxygen Abundances}
\tablehead{
\colhead{Star} &  
\colhead{[\ion{Fe}{2}/H]} &
\colhead{[C/\ion{Fe}{2}]} &
\colhead{[O/\ion{Fe}{2}]}  
}
\startdata
HD 2796    & $-$2.26 & $-$1.00 & +0.57  \\
HD 6268    & $-$2.35 & $-$1.13 & +0.53   \\
BD-180271  & $-$2.35 & $-$0.73 & +0.56 \\
HD 110184  & $-$2.41 & $-$0.58 & +0.51 \\ 
HD 122956  & $-$1.60 & $-$0.50$^a$ & +0.30 \\
HD 134440  & $-$1.44 &  +0.19$^b$ & +0.31 \\
BD-185550  & $-$2.87 & $-$0.39 & +0.53 \\
\enddata
\tablecomments{(a) From 1.56 $\micron$ CO lines;
               (b) Adopted from Carbon et al. 1987}
\label{tabxxxxx}
\end{deluxetable}

\begin{deluxetable}{cccccc}
\tablewidth{0pt}
\footnotesize
\tablecaption{Sensitivity to  \teff, log {\it g} and $v_t$  for HD 110184}
\tablehead{
\colhead{Abundance}  &
\colhead{$\Delta$ \teff} &
\colhead{$\Delta$ log $g$} &
\colhead{$\Delta v_t$} &
\colhead{($\Sigma$x$^2$)$^{1/2}$} \\
\colhead{}  &
\colhead{+100 K} &
\colhead{+0.5 dex} &
\colhead{+0.5 km s$^{-1}$} &
\colhead{}
}
\startdata
{[FeI/H]}  & +0.15   & -0.18   & $-$0.09 & 0.25 \\
{[FeII/H]} & $-$0.04 & +0.13   & $-$0.06 & 0.15 \\
{[O/H]}    & +0.22   & $-$0.24 & +0.04   & 0.33 \\
{[O/FeI]}  & +0.07   & $-$0.06 & +0.13   & 0.16 \\
{[O/FeII]} & +0.26   & $-$0.37 & +0.10   & 0.46 \\
\enddata
\label{tabxxxxx}
\end{deluxetable}

\begin{deluxetable}{llllll}
\tablewidth{0pt}
\footnotesize
\tablecaption{Mean [O/Fe] from [\ion{O}{1}] in bins of 0.2 dex in [Fe/H]}
\tablehead{
\colhead{[Fe/H]}    &
\colhead{[O/Fe]}    &
\colhead{$\sigma$}  &
\colhead{$\sigma_{mean}$} &
\colhead{Stars}
}
\startdata
0.0 & $-$0.04 & 0.13 & 0.04 & 11 \\
$-$0.2 & 0.06 & 0.07 & 0.02 & 11 \\
$-$0.4 & 0.12 & 0.08 & 0.02 & 16 \\
$-$0.6 & 0.19 & 0.09 & 0.02 & 13 \\
$-$0.8 & 0.23 & 0.13 & 0.03 & 14 \\
$-$1.0 & 0.25 & 0.14 & 0.07 & 4 \\
$-$1.2 & 0.33 & 0.10 & 0.03 & 12 \\
$-$1.4 & 0.37 & 0.08 & 0.02 & 15 \\
$-$1.6 & 0.36 & 0.10 & 0.02 & 19 \\
$-$1.8 & 0.34 & 0.12 & 0.03 & 14 \\
$-$2.0 & 0.37 & 0.10 & 0.04 & 5 \\
$-$2.2 & 0.42 & 0.13 & 0.05 & 7 \\
$-$2.4 & 0.40 & 0.08 & 0.03 & 7 \\
$-$2.6 & 0.47 & 0.18 & 0.08 & 5 \\
$-$2.8 & 0.53 & 0.04 & 0.03 & 2 \\
$-$3.0 & 0.66 & --  & --  & 1 \\
\enddata
\label{bin}
\end{deluxetable}

\begin{figure}
\epsscale{}
\plotone{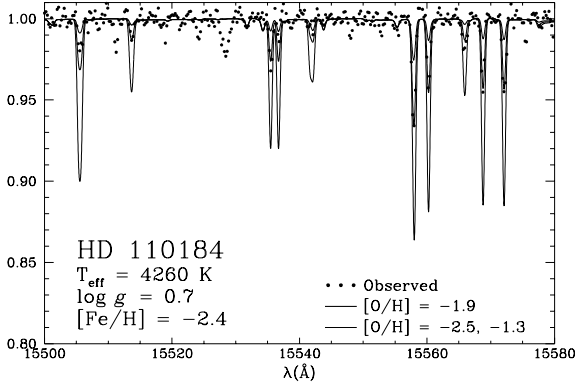}
\caption{Spectrum of HD 110184 ({\it circles}) compared with synthetic spectra 
computed with  [O/H]: -1.9 ({\it thick line}) and -2.5 and -1.3 ({\it thin lines})}
\label{hd110184OH}
\end{figure}

\begin{figure}
\epsscale{}
\plotone{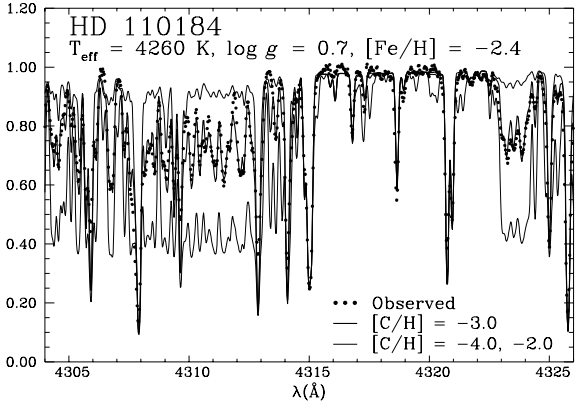}
\caption{Spectrum of HD 110184 ({\it circles}) compared with synthetic spectra 
computed with  [C/H]: -3.0 ({\it thick line}) and -4.0 and -2.0 ({\it thin lines})}
\label{hd110184CH}
\end{figure}

\begin{figure}
\epsscale{}
\plotone{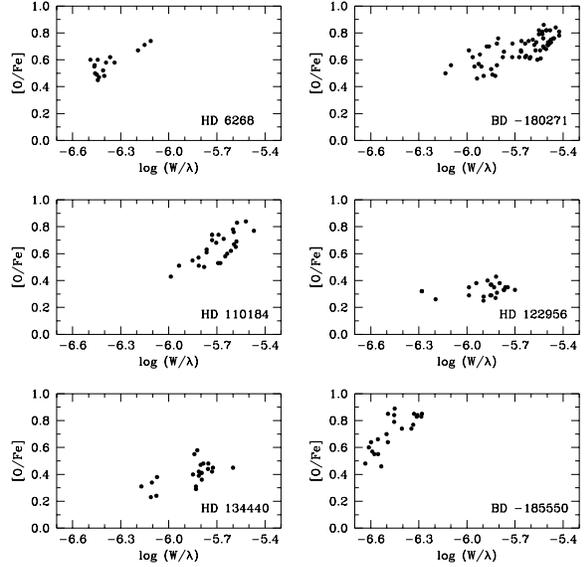}
\caption{[O/Fe] vs. reduced equivalent widths for IR OH lines (see Table 7). 
}
\label{ofeew}
\end{figure}

\begin{figure}
\epsscale{}
\plotone{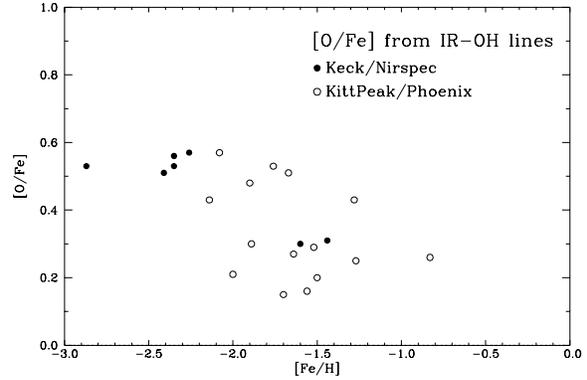}
\caption{[O/Fe] vs. [Fe/H] from IR OH lines. Filled circles show results
from the Keck NIRSPEC spectra (this work), and open circles show results
from revised Phoenix data (see Mel\'endez et al. 2001 and Table 5). 
}
\label{ofeab}
\end{figure}

\begin{figure}
\epsscale{}
\plotone{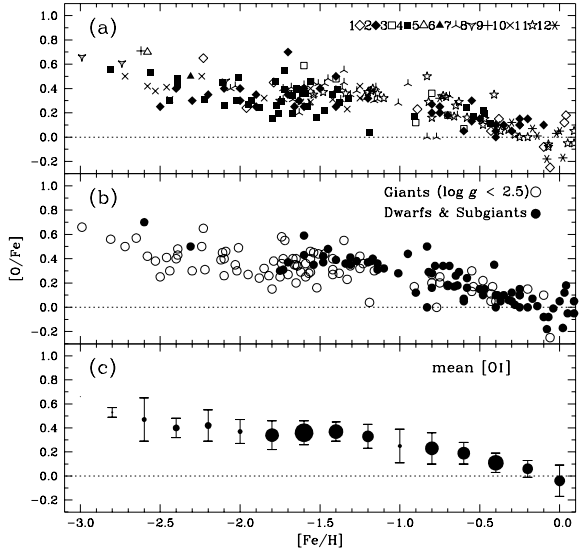}
\caption{[O/Fe] vs. [Fe/H] from [\ion{O}{1}] lines. {\it Top}: Results
from (1) Gratton \& Ortolani 1986; 
(2) Barbuy 1988, Barbuy \& Erdelyi-Mendes 1989; (3) Spite \& Spite 1991;
(4) Sneden et al. 1991, Kraft et al. 1992, Shetrone 1996;
(5) Balachandran \& Carney 1996, Balachandran et al. 2001;
(6) Fulbright \& Kraft 1999; (7) Gratton et al. 2000;
(8) Westin et al. 2000; Cowan et al. 2002; (9) Cayrel et al. 2001;
(10) Sneden \& Primas 2001; 
(11) Nissen et al. 2001, Nissen \& Edvardsson 1992;
(12) Smith et al. 2001.
{\it Middle}: Giants ({\it open circles}), log {\it g} $<$ 2.5;
Dwarfs and subgiants ({\it filled circles}), log {\it g} $\geq$ 2.5.
{\it Bottom}: Mean values and $\sigma$ from Table 10; the size of the circles 
represents the number of stars in each bin of 0.2 dex in [Fe/H].
}
\label{ofeab}
\end{figure}



\end{document}